\documentclass[useAMS,usenatbib]{mn2e}
\usepackage{graphicx}
\usepackage{txfonts}
\title[INTEGRAL observations of AGNs obscured by the Galactic Plane]
  {INTEGRAL observations of AGNs obscured by the Galactic Plane}
\author[M. Molina et al.]
{M.~Molina,$^1$
  A.~Malizia,$^2$ L.~Bassani,$^2$ A.J.~Bird,$^1$
  A.J.~Dean,$^1$ R.~Landi,$^2$ A.~De Rosa,$^3$ 
\newauthor
 R.~Walter,$^4$ E.J.~Barlow,$^1$ D.J.~Clark,$^1$ A.B.~Hill,$^1$ V.~Sguera$^1$\\
 $^1$School of Physics and Astronomy, University of Southampton,
        SO17 1BJ, Southampton, U.K.,\\
 $^2$IASF/INAF, via Gobetti 101, I-40129 Bologna, Italy,\\
 $^3$IASF/INAF, via del Fosso del Cavaliere 100, I-00133 Roma, Italy,\\
 $^4$INTEGRAL Science Data Centre, Chemin d'Ecogia 16, CH-1290 Versoix, Switzerland.}
\begin{document}

\date{}

\pagerange{\pageref{firstpage}--\pageref{lastpage}} \pubyear{2006}

\maketitle

\label{firstpage}
        
\begin{abstract}

In this paper we present INTEGRAL observations of 7 AGNs: two newly discovered type 1 Seyferts,
IGR J18027-1455 and IGR J21247+5058, and five well known Seyferts, NGC 6814 (type 1.5), Cyg A
(Type 2), MCG-05-23-16 (type 2), ESO 103-G035 (type2) and GRS1734-292. For IGR J18027-1455 and
IGR J21247+5058 only INTEGRAL/IBIS data were available, while broadband spectra are presented
and discussed for the remaining 5 sources for which either BeppoSAX or ASCA data 
were used in conjunction with INTEGRAL measurements. In the cases of NGC 6814 and GRS 
1734-292, data taken in different periods indicate variability in the flux: in the case of NGC 
6814 by a factor of 16 over a period of about 10 years.
Although limited in size, our sample can be used to investigate the parameter space of both 
the photon index and cut-off energy. The mean photon index is 1.8, while the cut-off energy
ranges from 30-50 keV to greater than 200 keV; in the particular case of MCG-05-23-16, ESO 
103-G035 and GRS 1734-292 the cut-off energy is well constrained at or below 100 keV.
We have also tested an  enlarged sample, which includes INTEGRAL data of 3 more AGNs,
against the correlation found by a number of authors between the 
photon index and the cut-off energy but have found no evidence for a relation between these two 
parameters. Our analysis indicates that there is a diversity in cut-off 
energies in the primary continuum of Seyfert galaxies. 
 
\end{abstract}


\section{Introduction}
The high energy emission of AGNs (Active Galactic Nuclei) is often to the first
order well described by a power law of photon index 1.8-2.0, extending from a few keV to
over 100 keV; at higher energies there is evidence of an exponential cut-off, the
exact value of which is still uncertain \citep{b25, b29, b41}. 
Secondary features, which are also commonly present, are considered to be the effects of reprocessing of
this primary continuum and are relatively well understood  \citep{b42, b41}. 
Modelling of high energy AGNs spectra has so far generally focussed on how to reproduce and
explain the observed primary continuum shape. A good fraction of the
proposed models ascribe the power law to the inverse Compton
scattering of soft photons in a bath of {\textquotedblleft{hot electrons}\textquotedblright} (see for example \citealt{b43}).  
Variations to this baseline model depend on the energy distribution of these
electrons and their location in relation to the accretion disc.
Measuring both the primary continuum and its cut-off energy is
therefore crucial for understanding models and discriminating between them. While
the photon index distribution has been well investigated \citep{b22},
observational results on the cut-off energy have so far been limited
by the scarcity of measurements above 10-20 keV, with most
information coming from BeppoSAX broadband spectra. Analysis of type
1 and 2 AGNs \citep{b25} provides evidence for a
wide range of values in the cut-off energy, spanning from 30 to 300
keV and further suggests a possible trend of increasing cut-off energy
as the power law index increases; it is not clear however if this
effect is due to limitations in the spectral analysis or if it is
intrinsic to the sampled source populations. 

Direct INTEGRAL measurements enable us to obtain further observational results on the
primary continuum and high energy cut-off thus providing more refined
parameters for AGNs modelling. The sample of 12 AGNs first detected by
INTEGRAL \citep{b2} can be taken as a case study: they are
representative of the larger sample of AGNs recently reported in the
20-100 keV band \citep{b4} and are detected with adequate
statistics to allow more in depth analysis. In particular, they are
ideal objects for the study of the primary power law component, the
presence of any high energy cut-off and the existence of a relation
(if any) between these two parameters. Spectral analysis has so far
been reported for 5 Seyfert galaxies listed in this initial sample, 4
of type 2 (NGC 4945, Centaurus A, Circinus Galaxy and NGC 6300, \citealt{b32}) 
and one of type 1 (GRS 1734-292, \citealt{b30}) as
well as for the Blazar PKS 1830-211 \citep{b11}. In the
present paper, we concentrate on the remaining 6 objects: the two
newly discovered AGNs, IGR J18027-1455 and IGR J21247+5058 both of type
1, and the four known Seyferts NGC 6814 (type 1.5), Cygnus A (type 2),
MCG-05-23-16 (type 2) and ESO 103-G035 (type 2). The significance
of the INTEGRAL detection for the last two objects was quite
low at the time of the paper by \citet{b32}, but they
were analised anyway. Their analysis has therefore
been updated here as more exposure is now available.
INTEGRAL/IBIS data alone are reported for IGR J18027-1455 and IGR
J21247+5058 while broadband spectra are presented and discussed in
the other 4 cases since existing BeppoSAX and ASCA data are available
for use in conjunction with the INTEGRAL spectra. We have also
updated and reanalysed the data of the remaining objects in the
\citet{b2} sample but only in one case, that of GRS 1734-292, are the
results of our spectral analysis different from already published
values; analysis of the broadband ASCA/INTEGRAL spectrum of this object
is therefore re-discussed in the present study.
     
\begin{small}
\begin{figure}
\centering
\includegraphics[width=0.8\linewidth]{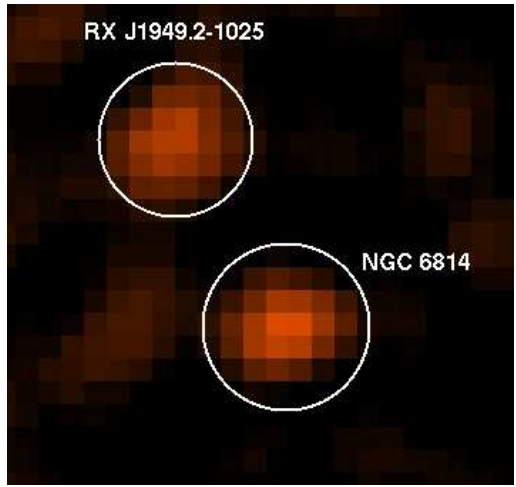}
\caption{ISGRI significance map of the region surrounding NGC 6814 and  
RX J1940.2-1025}
\label{fig1}
\end{figure}
\end{small}

\section{Data Analysis}
In this work we analyse the data of the imager IBIS \citep{b35} on 
board INTEGRAL \citep{b38}. 
This coded mask instrument is made by the combination of two detector layers: ISGRI 
\citep{b18}, an upper CdTe layer sensitive in the range between 15 keV and 1 MeV, and PICsIT 
\citep{b12}, a bottom CsI layer sensitive in the range 200 keV to 8 MeV;
in the present paper, we refer to data collected by the first layer
only, since the sources are too weak above $\sim$ 200 keV for detection by PICsIT.
The data reported here belong to the Core Programme (i.e. data
collected as part of the INTEGRAL Galactic Plane Survey and Galactic
Centre Deep Exposure \citep{b38}) as well as to public Open Time
observations, and span from revolution 46 (February 2003) to
revolution 234 (September 2004) included. A detailed description of the
source extraction criteria can be found in \citet{b5,b6};
briefly, ISGRI images for each available pointing are generated in  narrow energy bands using 
the INTEGRAL Science Data Centre (ISDC) offline scientific analysis software OSA version 4.2
\citep{b16}, including background uniformity corrections
\citep{b34}. Source ghosts are removed from each image using
a catalogue of sources built iteratively and containing at the end
all the detected objects. The clean images are then mosaiced using a custom
tool to produce deep all-sky maps in narrow energy bands. Images from adjacent 
bands can then be added together to estimate the source overall detection in a broad energy 
range; typically the 20-100 keV energy range is the
most suitable for detecting extragalactic objects \citep{b2}. Light curves are also created in the same narrow bands 
and then used for variability 
and/or spectral studies; in particular, the weighted mean value and associated error in each 
narrow energy band is calculated and then used to create a raw spectrum.

In Table 1 we report the details of the IBIS/ISGRI observations:
best fit positions, exposures and count rates  in the range 20-150 keV as well as the value of 
galactic absorption in each source direction  and the fluxes in the 20-100 keV band. 
All objects are detected with a global significance greater than 7$\sigma$; 
the positional uncertainty for sources of this brightness is around $\sim$1 arcmin \citep{b6}. 

Given the angular separation of $\sim$37' between NGC 6814 and the cataclysmic 
variable RX J1940.2-1025, previous X-ray data of this AGN, obtained with non-imaging collimated 
instruments, were generally contaminated. In particular, a past single observation at
high energy by OSSE \citep{b40} is clearly incorrect due to the inability of the 
instrument to resolve NGC 6814 from the Galactic object.
Thanks to the imaging capability of IBIS, the detection of NGC 6814 reported in this paper is 
the first reliable one above 10 keV. In figure \ref{fig1}, the IBIS/ISGRI 20-100 keV map of the 
sky region around NGC 6814 and RX J1940.2-1025 is shown; this figure indicates the ability of 
INTEGRAL to discriminate between the AGNs and the Galactic source.

\begin{table*}
\begin{center}
\centerline{{\bf Table 1: Observations Log}}
\vspace{0.5cm}
\begin{tabular}{lcccccc}
\hline
\hline
Source               &   RA    &  Dec     & Exposure& Count Rate$^{\dagger}$  & N$_{H}\ddagger$ & F(20-100 keV)\\
                     &         &          & (ksec)  & Counts/s            & 10$^{22}$ cm$^{-2}$ & 10$^{-11}$ erg cm$^{-2}$ s$^{-1}$\\
\hline
NGC 6814             & 295.666 & -10.329  &  252    & 0.69$\pm$0.06       &  0.13   & 5.8  \\
Cygnus A             & 299.869 & +40.733  &  426    & 1.02$\pm$0.05       &  0.35   & 17  \\  
IGR J18027-1455      & 270.690 & -14.922  & 1476    & 0.54$\pm$0.03       &  0.50   & 4.8  \\
IGR J21247+5058      & 321.151 & +50.980  &  438    & 1.21$\pm$0.04       &  1.12   & 10.53  \\
MCG-05-23-16         & 146.985 & -30.932  &  337    & 2.68$\pm$0.05       &  0.08   & 16.6  \\
ESO103-G35           & 279.695 & -65.408  &   41    & 1.08$\pm$0.15       &  0.08   & 7.4  \\
GRS1734-292          & 264.369 & -29.140  & 4040    & 1.08$\pm$0.01       &  0.76   & 7.3   \\
\hline
\hline
\end{tabular}
\end{center}
\small 
$\dagger$: in the 20-150 keV band; $\ddagger$: Galactic column density from \citealt{b41} 
\end{table*}

\begin{table*}
\begin{center}
\centerline{{\bf Table 2: Observations Periods}}
\vspace{0.5cm}
\begin{tabular}{lccc}
\hline
\hline
Name            &  IBIS/ISGRI         &  BeppoSAX NFI & ASCA/GIS \\
\hline
NGC 6814        & Feb. 2003/Sep. 2004 &        -      & May 1993 \\
Cygnus A        & Feb. 2003/Sep. 2004 &  Oct. 1999    &   -      \\
IGR J18027-1455 & Feb. 2003/Sep. 2004 &        -      &   -      \\
IGR J21247+5058 & Feb. 2003/Sep. 2004 &        -      &   -      \\
MCG-05-23-16    & Feb. 2003/Sep. 2004 &  Apr. 1998    &   -      \\
ESO103-G35      & Feb. 2003/Sep. 2004 &  Oct. 1996    &   -      \\
GRS1734-292     & Feb. 2003/Sep. 2004 &        -      & Sep. 1998\\
\hline
\hline
\end{tabular}
\end{center}
\small 
\end{table*}

To compile broadband spectra of our sources we either use ASCA/GIS or BeppoSAX/MECS+PDS 
data available in their respective archives. See Table 2 for a listing of the various observation periods.
The ASCA data, spectra and associated 
files were downloaded from the TARTARUS database (Version 3.1, {\it 
http://tartarus.gsfc.nasa.gov)} or from the HEASARC archive {\it 
(http://heasarc.gsfc.nasa.gov)}; the BeppoSAX/MECS data were 
downloaded from the ASDC (ASI Science Data Center) archive {\it 
(http://www.asdc.asi.it/bepposax/)}. 
The PDS spectra were extracted using the XAS package \citep{b9} in 
order to perform an in depth check for possible contaminations in the background fields,
which is an effect likely to occur in sources located in the Galactic Plane (see 
\citealt{b32}).

The spectral analysis was performed using XSPEC version v.11.3.1 \citep{b1}. Errors are quoted 
at the 90\% confidence level for the model parameter considered ($\Delta\chi^2$=2.7). In the 
following, we always use an absorbed component to take into account the Galactic
column density {\bf (wa$_g$)}, which in the direction of all our objects is significant due to their 
low Galactic latitude. Extra absorption intrinsic to the source {\bf (wa)}
was assumed when required by the data or known to be present from the literature.
To account for possible cross-calibration mismatches between instruments as well as to take 
into account flux variations between the observing periods,
a constant factor {\bf(C)} has always been added to the fit; this constant has been fixed to be 1 
between the two ASCA-GIS instruments while it  has been allowed to vary within the nominal 
range of values (0.75-0.95) in the case of the two BeppoSAX instruments (MECS versus PDS, 
\citealt{b15}); only with respect  to the INTEGRAL/ISGRI data,
has the constant been allowed to vary freely.

\section{Previously unpublished sources}
In this section we present results related to 4 objects for which INTEGRAL spectral  data
have not yet been reported nor discussed in the literature: the well known AGNs, NGC 6814 and 
Cygnus A, and two newly discovered galaxies, IGR J18027-1455 and IGR J21247+5058.
The baseline model used for these sources is a simple power law seen through galactic absorption
(\texttt{wa$_g$*po} in XSPEC); when required extra absorption (\texttt{wa}) and a narrow gaussian line (\texttt{ga}) have been added to this baseline model.

\subsection{NGC 6814}
The combined ASCA/GIS and INTEGRAL/ISGRI data (2-150 keV) are well fitted ($\chi^{2}$=91.7/106) by a single power law having a photon index of 
$\Gamma$$\sim$1.6 plus a cold iron line (see Table 3). 
The broadband spectrum obtained with this model is shown in figure \ref{ngc6814}. The iron line is strongly required by the data ($\Delta\chi^2$=11.2 for 2 d.o.f., i.e. more than 99\% significant) and has an equivalent width of  $\sim$400 eV, higher but compatible within errors with reflection in the accretion disk. 
Addition of a reflection component having R=1 (where R is the solid angle in units of 2$\pi$
subtended by the reflecting material \citep{b43}) does not improve the fit, probably due to the 
statistical quality of the data which is not sufficient to constrain this component. 
Similarly, the introduction of a cut-off to the power law (\texttt{cutoffpl} model in XSPEC)
does not improve the quality of the fit but allows us to put a lower limit at E$_{cut}>$70 keV. 
NGC 6814 is a Seyfert galaxy which shows strong variability in its X-ray flux over short 
(hours, \citealt{b33}) as well as long (days and years, \citealt{b24}) timescales. 
The constant \emph{C}, introduced to take into account a possible cross-calibration mismatch between the two instruments and/or flux changes between the two observations, gives a value of 
16$^{+9}_{-5}$. Since from previous works \citep{b36, b20} we know that the 
cross-calibration constant between ASCA/GIS and INTEGRAL/ISGRI is close to 1, we can 
conclude that, as expected, there was a strong variation between the ASCA and INTEGRAL  
observations, with INTEGRAL providing a higher flux than ASCA.
It is worth noting that since we are dealing with
the first high energy data not contaminated by the nearby Galactic source, this flux variation 
is completely due to the Seyfert galaxy.
The constant versus photon index contours obtained using our baseline model are shown in 
figure \ref{contours}. 
The ASCA 2-10 keV flux of 1.8 $\times$ 10$^{-12}$ erg cm$^{-2}$s$^{-1}$ and its extrapolated
20-100 keV flux of about 4$\times$10$^{-12}$erg cm$^{-2}$s$^{-1}$ suggest that the 
source was in a very low state during 1993 but had returned to the high state for the
first observations with INTEGRAL in May 2003 (see Table 1), much in agreement to what was observed by XTE 
(see also figure 5 in \citealt{b24}).
While INTEGRAL observations confirm and extend to the high energy domain the presence of long 
term variability, flux changes on shorter timescales are more difficult to assess due 
to the lower statistical significance of the signal. No changes are evident in the ISGRI 
light curve over the entire observing period.

Similarly to the X-ray band, strong variations have been observed in the optical continuum and 
line emission on timescales of months and longer \citep{b14,b31}: the width of the broad 
emission lines varies to the extent that the source classification changes in time from Seyfert 
1 to Seyfert 1.8/1.9. 
Furthermore, the rapid decrease in the broad emission line strengths is thought to be coupled 
with the observed X-ray flux behaviour \citep{b17,b24}. 
For its extreme variability and also for the modest width of the optical emission lines, Mukai 
and co-workers (2003) suggest that NGC 6814 might belong to the class of Narrow Line Seyfert 1 
(NLS1s) galaxies. 
However, contrary to what we find here, NLS1s have X-ray continua characterised by steeper 
slopes in the soft and hard X-ray bands \citep{b7,b8} than their
broad line counterparts. The soft excess is another characteristic of this class of
AGNs, but there is no evidence for this feature in our data (the 0.5-2 keV band has
not been reported here due to the low statistics); furthermore the primary 
power law component is harder than generally observed in NLS1 galaxies. Clearly a better 
understanding of this peculiar object can only come from coordinated optical/high energy 
observations with improved temporal coverage. 
Owing to its location close to the Galactic plane, it is likely that INTEGRAL will continue
serendipitous monitoring of this source and so provide further insight into its variability 
behaviour.

\begin{table*}
\begin{center}
\centerline{{\bf Table 3: Spectral Best-Fits for the Four Unpublished Sources}}
\vspace{0.5cm}
\begin{tabular}{lcccccc}
\hline
\hline
Source    & N$_H^\dagger$       & $\Gamma$              & E$_{line}$           & EW                 & F(2-10 keV) & $\chi^2$(d.o.f) \\
          &                     &                       &  keV                  & eV                 & 10$^{-11}$erg cm$^{-2}$s$^{-1}$ &                 \\
\hline
\hline
NGC6814  &   -                  & 1.67$^{+0.18}_{-0.11}$& 6.35$^{+0.15}_{-0.23}$& 472$^{+235}_{-228}$ &    0.18      & 91.7 (106)            \\
Cygnus A & 27.4$^{+2.3}_{-2.2}$ & 1.78$^{+0.01}_{-0.02}$&    -                 &           -          &   6.0       & 100.5 (83)      \\  
IGR J18027-1455 &  -            & 2.13$^{+0.23}_{-0.22}$&    -                 &           -          &    -         & 2.54   (4)          \\  
IGR J21247+5058 &  -            & 1.87$^{+0.21}_{-0.19}$&    -                 &           -          &   -         & 4.0   (4)             \\  
\hline
\hline
\end{tabular}
\end{center}
\small
$\dagger$ intrinsic column density
\end{table*}

\begin{small}
\begin{figure}
\centering
\includegraphics[width=0.6\linewidth,angle=-90]{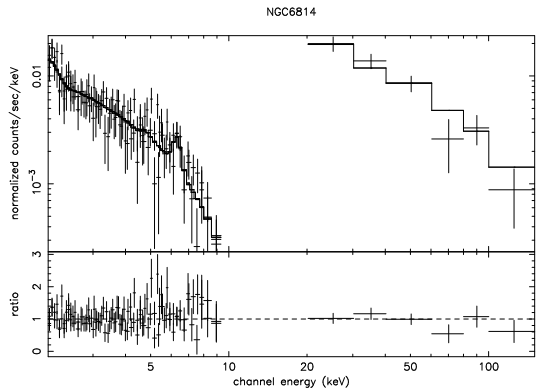}
\caption{INTEGRAL/ISGRI and ASCA/GIS broadband spectrum of NGC 6814: the model is a power law 
absorbed by the Galactic column density plus a cold iron line.}
\label{ngc6814}
\end{figure}

\end{small}
\begin{small}
\begin{figure}
\centering
\includegraphics[width=0.6\linewidth,angle=-90]{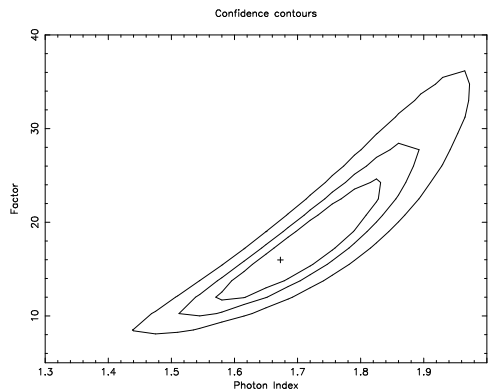}
\caption{Photon index of the primary power law against the INTEGRAL/ISGRI-ASCA/GIS cross 
calibration constant \emph{C}, indicating strong variability in flux between the two satellites 
measurement epochs.}
\label{contours}
\end{figure}
\end{small}

\subsection{Cygnus A}
The radio galaxy Cygnus A is an active galaxy at the centre of a cooling flow in a 
cluster of galaxies \citep{b28}. Due to this unfavourable location, the 
determination of its high energy spectrum is not easy.
Hard X-ray emission from Cygnus A has already been reported by \citet{b39} with the 
RXTE/HEXTE instrument and a previous detection by the PDS instrument on 
board BeppoSAX is available but was never published. In the present work, 
the BeppoSAX (MECS-PDS) and IBIS-ISGRI data are combined and analysed with the aim of defining 
the nuclear spectrum of Cygnus A.
To consider both the cluster and the active nucleus, the overall data have been fitted
with a bremsstrahlung component plus an intrinsically absorbed power law component.
The thermal component refers to the cluster emission while the power law component takes 
into account the AGN emission; both components
are seen though Galactic absorption.
Moreover, from the 2-10 keV residuals a strong excess is evident above 6 keV, 
therefore a Gaussian line has been added to the bremsstrahlung component to take into account 
the cluster gas iron line. 
Therefore the overall model is (\texttt{wa$_g$*(bremss+ ga + wa*(po)}).
Fixing the cluster gas temperature to the Chandra value of kT=6.8 
keV \citep{b39}, the model provides our best fit and even when this value is 
left free to vary we find it compatible with the Chandra estimates. Overall, the physical 
parameters of both components are in agreement with previous observations \citep{b39}: 
the cluster gas has an iron line at 6.78$^{+0.02}_{-0.05}$ keV
with an equivalent width of 487 eV, while the AGN has a photon index $\Gamma$ 
of 1.8 absorbed by a column density of N$_{H}$$\sim$3$\times$10$^{23}$cm$^{-2}$,
compatible with the Seyfert 2 nature of Cygnus A (see Table 3).   
In figure \ref{cygnA} the combined BeppoSAX/INTEGRAL broadband 2-150 keV spectrum is shown.
To this best fit model, we also tried the addition of an extra Gaussian line to account for the 
AGNs iron line emission, but this feature is not required by the data; the upper limit on the 
equivalent width of any nuclear iron line is $\sim$1 keV.
Also to account for a possible cross-calibration mismatch between the two measurements 
(BeppoSAX and INTEGRAL) as well as for flux variations between the two observing periods,
a constant factor C has been added to the fit; when left free to vary it provides values in the 
range 0.93-1.27, implying a good match between the two observations and therefore no evidence 
for strong  variability. Finally, we substituted the simple power law with a cut-off power law; 
the fit does not improve but provides a lower limit to any spectral drop off in the source at 
E$_cut$$>$250 keV. Thanks to the present analysis we are able to estimate the contribution of the 
cluster as well as that of the AGNs at energies $>$20 keV: if we consider only the cluster, the 
ISGRI (PDS) flux is F$_{20-100 keV}$ = 2.4 (2.1) $\times$ 10$^{-12}$ erg cm$^{-2}$ s$^{-1}$, 
which means that the high energy emission listed in Table 1 is completely dominated by the active nucleus of 
Cygnus A. 

\begin{small}
\begin{figure}
\centering
\includegraphics[width=0.6\linewidth,angle=-90]{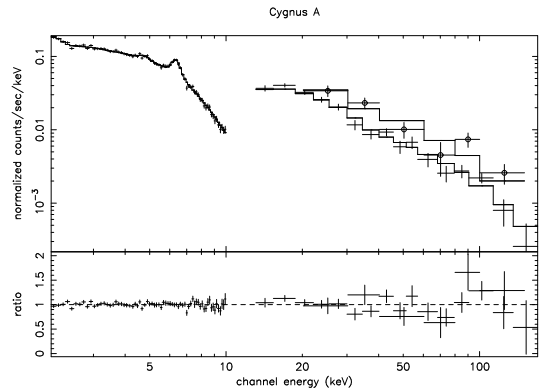}
\caption{INTEGRAL/ISGRI and BeppoSAX/MECS-PDS broad band spectrum of Cygnus A: the model 
contains a galaxy cluster component (bremsstrahlung plus iron line) and AGNs component (absorbed 
power law affected both by intrinsic and Galactic absorption in the source direction.}
\label{cygnA}
\end{figure}
\end{small}

\subsection{IGR J18027-1455 and IGR J21247+5058}
IGR J18027-1455 and IGR J21247+5058 are totally new discoveries, and although they were soon 
recognised as AGN candidates, only recently they have been identified with active galaxies
\citep{b21}. IGR J18027-1455 is a Seyfert of type 1 at a redshift z=0.035. IGR 
J21247+5058, instead, is a radio source with a quite puzzling optical spectroscopic appearance 
showing a broad, redshifted H$\alpha$ complex at z=0.020
superimposed onto a {\textquotedblleft{normal}\textquotedblright} F/G-type Galactic star 
continuum. These features, together with the spatially coincident extended radio emission, 
suggest a chance alignment between a relatively nearby star and a background radio galaxy.
Since no data at low energy are available for either object, ISGRI data have been 
simply fitted using an absorbed power law with the column density fixed to the galactic value.  

This model provides a good fit to the spectrum of both sources and a photon index close to 2 
e.g. typical of active galaxies. The source spectra and residuals from this model
are shown in figure \ref{IGRJ18027-1455} and \ref{IGRJ21247}, while the best fit parameters are 
reported in Table 3. We have also searched for possible high energy cut-offs in the spectrum of 
both galaxies (again using the \texttt{cutoffpl} model)
but again only lower limits  have been found which are E$_{cut}$$>$40 keV and E$_{cut}$$>$30 keV
for IGR J18027-1455 and IGR J21247+5058 respectively. We also tried to fix the photon index to 
1.8 for both sources in order to better constrain the high energy cut-off, but while for 
IGR J18027-1455 we were able to find a range for E$_{cut}$ of 90-786 keV, for IGR J21247+5058 the
value of E$_{cut}$ is still not constrainable. 

\begin{small}
\begin{figure}
\centering
\includegraphics[width=0.6\linewidth, angle=-90]{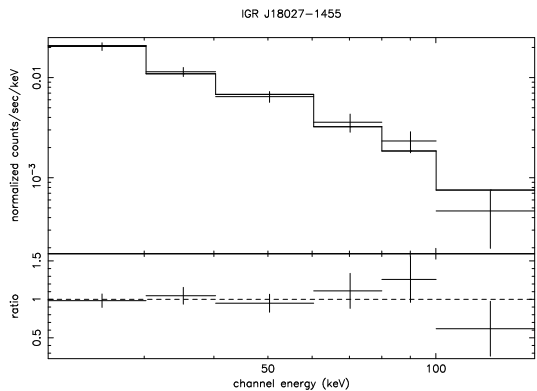}
\caption{ISGRI spectrum of IGR J18027-1455 and residuals: the model is an absorbed power law 
with column density fixed to the Galactic value.}
\label{IGRJ18027-1455}
\end{figure}
\end{small}

\begin{small}
\begin{figure}
\centering
\includegraphics[width=0.6\linewidth,angle=-90]{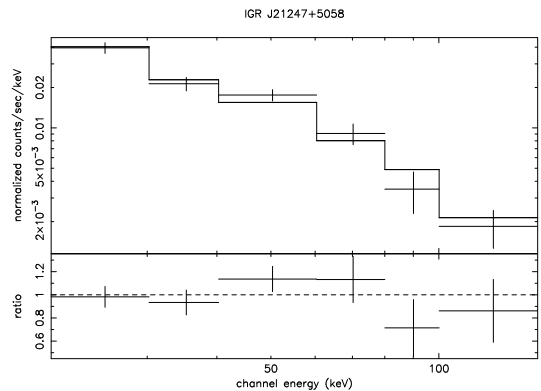}
\caption{ISGRI spectrum of IGR J21247+5058 and residuals: the model is an absorbed power law 
with column density fixed to the Galactic value.}
\label{IGRJ21247}
\end{figure}
\end{small}
  
\section{Previously published sources}
Next we concentrated on the re-analysis of INTEGRAL data of 3 more 
sources (MCG-05-23-16, ESO103-G035 and GRS 1734-292), extracted from the first sample of AGNs 
detected by INTEGRAL \citep{b2}. In the first two cases more exposure is now 
avalaible to allow a more in depth study and a good constraint of the primary emission 
characteristics (see also \citealt{b32} for a discussion of the BeppoSAX data).
In the case of GRS 1734-292 our analysis provides slightly different results than a 
previously published study \citep{b30}. 
The baseline model used in this section is an exponentially cut-off power law spectrum reflected 
from neutral material plus a narrow gaussian line; the Compton reflection
component is characterized by the parameter R  and components are seen 
through both galactic and intrinsic absorption  (\texttt{wa$_g$*wa*(pexrav+gauss)}).
We find that in all these 3 AGNs the primary 
continuum is well characterised and the high energy cut-off properly constrained.

\begin{table*}
\begin{center}
\centerline{{\bf Table 4: Spectral Best-Fits for the Previously Published Sources}}
\vspace{0.5cm}
\begin{tabular}{lcccccccc}
\hline
\hline
Source         & N$_H^\dagger$ & $\Gamma$              & E$_{line}$           & EW                 & $R^\sharp$ & E$_{cut}$ & F(2-10 keV)  & $\chi^2$(d.o.f) \\
               &                        &                       &  eV                  & eV                 &   &keV        & 10$^{-11}$erg cm$^{-2}$s$^{-1}$ &      \\
\hline
\hline
MCG-05-23-16   & 1.45$^{+0.19}_{-0.20}$ & 1.74$^{+0.08}_{-0.14}$ & 6.4$^{+0.07}_{-0.10}$  & 89$^{+20}_{-20}$ &1.2$^{+0.6}_{-1.0}$ & 112$^{+40}_{-43}$ & 9.3 & 82.1 (74) \\  
ESO103-G35     & 18.8$^{+2.16}_{-1.12}$ & 1.78$^{+0.16}_{-0.22}$ & 6.39$^{+0.07}_{-0.07}$ & 172$^{+80}_{-34}$ & $<$1.9  & 68$^{+71}_{-25}$ &   3.8 & 170.4 (174) \\        
GRS1734-292    & 1.05$^{+0.58}_{-0.64}$ & 1.74$^{+0.19}_{-0.22}$ &    -                 &           -        & 1 (fixed) & 58$^{+22}_{-14}$ & 3.7 & 131.5 (78)   \\  
\hline
\hline
\end{tabular}
\end{center}
\small
$\dagger$ intrinsic column density\\
$\sharp$ reflection parameter
\end{table*}

\subsection{MCG-05-23-16}
The broadband spectrum of this source has already been discussed by
\citet{b29} and \citet{b23} using the same
BeppoSAX data employed in this paper, with additional RXTE observations 
below 30 keV in the case of the Mattson and Weaver paper.  
However, only Risaliti takes into account the possible
presence of a cut-off in the high energy spectrum of this source so
that our results can be directly compared to his ones. Combining our
ISGRI data with the BeppoSAX points provides a good fit (see Table 4) and 
a set of parameters which are substantially in agreement with
those obtained by \citet{b29}, once the relative uncertainties  
and the use of a different inclination angle between the reflecting material and the line of sight 
(30$^{\circ}$ in Risaliti and 63$^{\circ}$ in the present work) are taken into account. 
Our choice of inclination is dictated by the classification of MCG-05-23-16 as a
Seyfert 2 galaxy, since this type of objects are more likely to be
seen edge-on. Fixing the angle to 30$^{\circ}$ provides an equally good
fit with parameters similar to those reported in Table 4, except for a slightly higher energy
cut-off energy (124 keV) and a stronger reflection (R=0.9). In both cases 
the reflection we obtain is higher than the one measured by \citet{b29} and \citet{b23}. 
To account for a possible
cross-calibration mismatch between ISGRI and MECS as well as for flux
variations between the two observing periods, the constant factor C has
been added to the fit; since the value obtained for this constant is close to 1, we
assume that no significant variation has occured in the source between BeppoSAX
and INTEGRAL observing periods. In this source the primary continuum is well constrained
both in terms of photon index and cut-off energy 
as evident in figure \ref{mcg_cont}, where one parameter is plotted  
against the other.

\begin{small}
\begin{figure}
\centering
\includegraphics[width=0.6\linewidth,angle=-90]{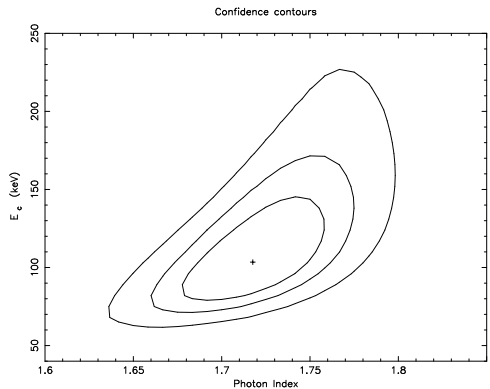}
\caption{Constraints on the primary continuum (power law photon index versus cut-off energy) of
MCG-5-23-16: contours are obtained for the model reported in Table 4.}
\label{mcg_cont}
\end{figure}
\end{small}

\subsection{ESO103-G035} 
For this galaxy two BeppoSAX observations are available but only one, performed on October 3, 
1996, is compatible in flux with the INTEGRAL measurement.
This BeppoSAX observation has been analysed by \citet{b37} and \citet{b29}. 
Even if the models used in these two papers are slightly different, the basic results 
are in agreement and the only noticeable discrepancy is on the high energy cut-off value: 
Risaliti locates the power law cut-off at 84 keV, an energy which is higher than that (29 keV)
found by \citet{b37}. Our results, obtained using the model described in section 4 (see Table 4 and figure \ref{eso_cont}) 
are in agreement with these previous studies and locate the cut-off energy between these two values 
confirming that in this object the exponential decline in the power law shape occurs below 100 keV. 
The cross-calibration constant C between BeppoSAX and INTEGRAL is 0.7$\pm$0.2 compatible with no substantial 
change in the source flux. 

\begin{small}
\begin{figure}
\centering
\includegraphics[width=0.6\linewidth,angle=-90]{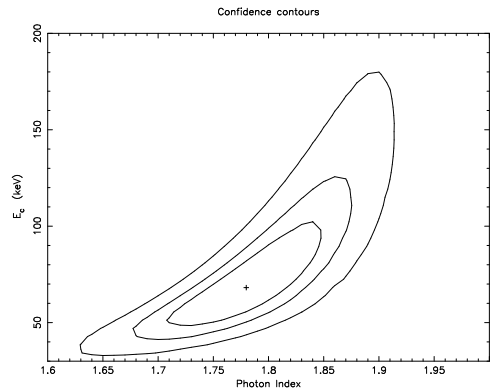}
\caption{Constraints on the primary continuum (power law photon index versus cut-off energy) of
ESO103-G035: contours are obtained for the model reported in Table 4.}
\label{eso_cont}
\end{figure}
\end{small}

\subsection{GRS 1734-292}
In this case ASCA/GIS and INTEGRAL/ISGRI data provide a broadband spectrum which is equally 
well fitted by an absorbed power law with no reflection (R=0) or with reflection R equal to 1
(see Table 3); in either case the primary power law component requires a cut-off at an energy  
below 100 keV (see figure \ref{grs_cont}). 
The fit with no reflection is slightly better ($\chi$$^{2}$=130.7 for 78 d.o.f.) and 
gives parameters ($\Gamma$=1.6, N$_H$=0.9 $\times$ 10$^{22}$ cm$^{-2}$ and E$_{cut}$=44 keV) 
which, within uncertainties, are compatible with those reported in Table 3. In either case,
the constant C introduced to account for instrument cross-calibration or variability is in the 
range 1.4-1.9, thus suggesting a change in the source intensity: a variation in flux of up to a 
factor of 2 is compatible with the X-ray time history of the source as reported in 
\citet{b30}. Our results are compatible with the best fit parameters obtained by 
these authors, except for the lower value of the cut-off energy which was poorly 
constrained to be above 100 keV in their analysis. If confirmed, the presence of a high energy 
cut-off at this value could put  the basis for another argument against the hypothesis that 
GRS1734-292 is a Blazar type of AGNs (see discussion in \citealt{b30}), and thus 
question its association with the 
unidentified EGRET gamma-ray source 3EG J1736-2908 \citep{b13,b30}.
In Blazars, in fact, the
cut-off energy is generally found at higher energies (in the MeV region), while we find a value below 100 keV. 
This cut-off energy is typical of Seyfert galaxies, which are well known to lack emission in the MeV domain 
\citep{b45}. 
It then follows that it is hard to match ASCA/INTEGRAL data with EGRET measurements if a cut-off is present 
at these low energies.

\begin{small}
\begin{figure}
\centering
\includegraphics[width=0.6\linewidth,angle=-90]{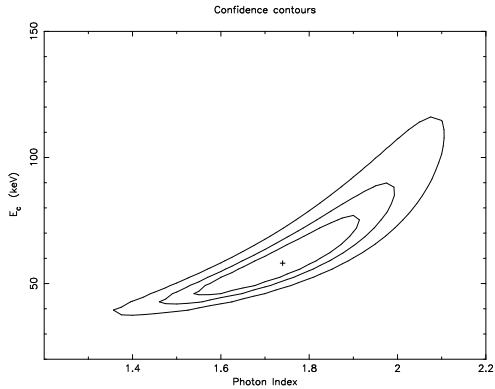}
\caption{Constraints on the primary continuum (power law photon index versus cut-off energy) 
of GRS1734-292: contours are obtained for the model reported in Table 4.}
\label{grs_cont}
\end{figure}
\end{small}

\section{Discussion and Conclusions}
Analysis of INTEGRAL data alone, or in conjunction with ASCA or BeppoSAX spectral data 
of a sample of 7 Seyfert galaxies of both type 1 and 2 provides information on the high energy 
emission of this class of objects. Only in the case of NGC 6814 and GRS 1734-292 do data taken 
in different periods indicate variability in the flux: in the case of NGC 6814 by a factor of 
16 over a period of about 10 years. 
Although limited in size, our sample can be used to investigate the parameter space of both 
$\Gamma$ and cut-off energy. The mean photon index is 1.83$\pm$0.07, while the cut-off energy
ranges from 30-50 keV to greater than 200 keV; in the case of MCG-05-23-16, ESO 
103-G035 and GRS 1734-292 the cut-off energy is well constrained at or below 100 keV.
We can also test our sample against the correlation found by a number of authors between the 
photon index and the cut-off energy. 
First claimed by \citet{b27}, it was further analysed and discussed by 
\citet{b22} and \citet{b26}, taking advantage of the broad band BeppoSAX 
observations of AGN. At present it is still debated, since previous works found that 
the two parameters are not independent in the fitting procedure, with $\Gamma$ increasing as 
the cut-off energy decreases \citep{b25}. In figure \ref{cutoff}, we plot the photon index 
versus the high energy cut-off for the sample of objects discussed here, to 
which we have also added the results obtained by \citealt{b32}. Although these authors only 
used high energy data above 10 keV to constrain the power law parameters, their results on 
$\Gamma$ and cut-off energy for NGC 4945, Circinus Galaxy and Centaurus A are quite solid being 
independently obtained by different instruments (see \citealt{b32} and references therein).  
Discussing this correlation in terms of the mechanisms that give rise to the main 
spectral component goes beyond the scope of this paper. We simply point out here 
the range of values that these two parameters occupy in our sample: this is an important 
observational information for modelling of the primary continuum but also for estimating the 
AGNs contribution to the X-ray Cosmic Diffuse Background \citep{b10}.
From figure \ref{cutoff} it is clear that our data do not show the correlation found by
\citet{b25}: the values for the photon index, in fact, cluster around 1.8, while the values of
the cut-off energy range over a wide interval. We can conclude that the correlation between the
high energy cut-off and the photon index remains to be proved and that a diversity in cut-off 
energy is most likely a property of Seyfert galaxies. 
INTEGRAL will keep observing AGNs for the duration of the mission and so we expect that more 
data on this class of objects will become available; this will provide further information on 
the high energy cut-off in AGNs and a deeper insight into the primary continuum emission.

\begin{small}
\begin{figure*}
\centering
\includegraphics[width=0.8\linewidth]{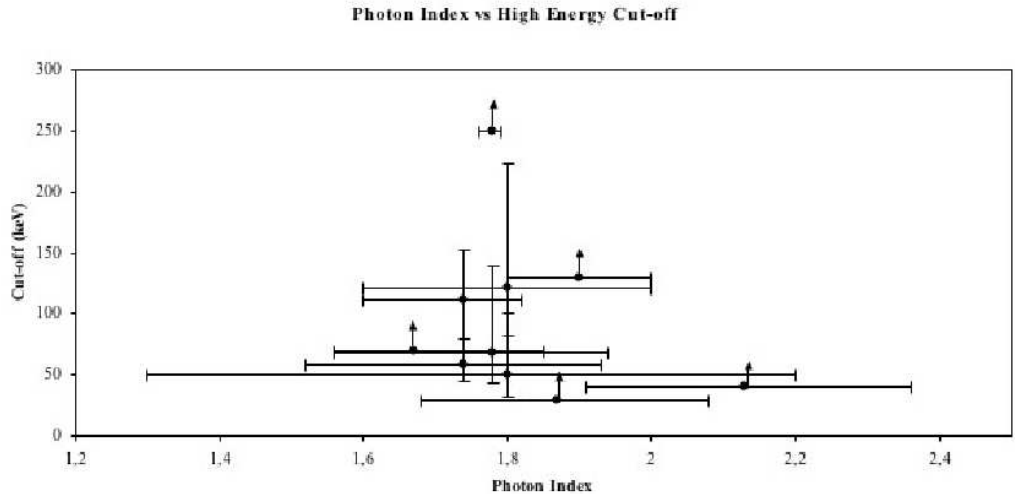}
\caption{Primary power law photon index against cut-off energy for the sample of AGNs analysed 
in the present paper plus 3 AGNs (NGC4945, Circinus galaxy and CenA) discussed by \citet{b32}.}
\label{cutoff}
\end{figure*}
\end{small}

\label{lastpage}

\newpage

\begin{thebibliography}{99}
\bibitem[\protect\citeauthoryear{Arnaud}{1996}]{b1} Arnaud K.A. 1996, Astronomical Data 
Analysis Software and Systems V,eds Jacoby G. and Barnes J., p.17, ASP Conf Series vol. 101
\bibitem[\protect\citeauthoryear{Bassani et al.}{2004}]{b2} Bassani L., Malizia A., Stephen 
J.B. et al. 2004 in Proceedings of the 5th INTEGRAL Workshop on the INTEGRAL Universe (ESA 
SP-552). 16-20 February 2004, Munich, Germany. Scientific Editors: V. Sch\"onfelder, G. Lichti 
\& C. Winkler, p.139
\bibitem[\protect\citeauthoryear{Bassani et al.}{2005}]{b3} Bassani L., De Rosa, A., Bazzano, 
A. et al. 2005, ApJ Lett., 634, 21
\bibitem[\protect\citeauthoryear{Bassani et al.}{2006}]{b4} Bassani L., Molina M., Malizia A. 
et al. 2006, ApJ Lett., 636, 65
\bibitem[\protect\citeauthoryear {Bianchi et al.}{2004}]{b41} Bianchi, S., Matt, G., Balestra, I.,Guainazzi, M., Perola, G. C. 2004, A\&A, 422, 65 
\bibitem[\protect\citeauthoryear{Bird et al.}{2004}]{b5} Bird A.J., Barlow E.J., Bassani L. et 
al. 2004, ApJ 607, 33
\bibitem[\protect\citeauthoryear{Bird et al.}{2006}]{b6} Bird A.J., Barlow E.J., Bassani L. et 
al. 2006, ApJ, 636, 765
\bibitem[\protect\citeauthoryear{Boller et al.}{1996}]{b7} Boller T., Brandt W. N., Fink H. 
1996, A\&A, 305, 53
\bibitem[\protect\citeauthoryear{Brandt et al.}{1997}]{b8} Brandt W. N., Mathur S., Elvis M. et 
al. 1997, AAS, 190, 5102 
\bibitem[\protect\citeauthoryear{Chiappetti \& Dal Fiume}{1997}]{b9} Chiappetti L., dal Fiume 
D. 1997 in Data Analysis in Astronomy, Proceedings of the Fifth Workshop. Ettore Majorana 
Centre for Scientific Culture, Erice, Italy. 27 Oct - 3 Nov, 1996. Edited by V. Di Gesu, M. J. 
B. Duff, A. Heck, M. C. Maccarone, L. Scarsi and H. U. Zimmerman. World Scientific Press, 
1997., p.10197, 777
\bibitem[\protect\citeauthoryear{Comastri et al.}{1995}]{b10} Comastri A., Setti G., Zamorani, 
G. et al. 1995, A\&A, 296, 1
\bibitem[\protect\citeauthoryear{De Rosa et al.}{2005}]{b11} De Rosa, A., Piro, L., Tramacere, 
A. et al. 2005, A\&A, 438, 121
\bibitem[\protect\citeauthoryear{Di Cocco et al.}{2003}]{b12} Di Cocco G., Caroli E., Malizia 
A. et al. 2003, A\&A 411, 189
\bibitem[\protect\citeauthoryear{Di Cocco et al.}{2004}]{b13} Di Cocco G., Foschini L., Grandi 
P. et al. 2004, A\&A, 425, 89
\bibitem[\protect\citeauthoryear{Dickey \& Lockman}{1990}]{b41} Dickey J.M. \& Lockman F.J. 
1990, ARA\&A, 28, 215
\bibitem[\protect\citeauthoryear{Doroshenko}{1988}]{b14} Doroshenko V.T. 1988, Astrofizika, 
28, 233
\bibitem[\protect\citeauthoryear{Fiore et al.}{1999}]{b15} Fiore, F., Guainazzi, M., \& Grandi, 
P. 1999, Handbook for BeppoSAX NFI spectral analysis, {\texttt 
ftp://ftp.asdc.asi.it/pub/sax/doc/software\_docs/saxabc\_v1.2.ps.gz}
\bibitem[\protect\citeauthoryear{Goldwurm et al.}{2003}]{b16} Goldwurm A., David P., Foschini 
L. et al. 2003, A\&A, 411, 223
\bibitem[\protect\citeauthoryear{K\"onig et al.}{1997}]{b17} K\"onig M., Friedrich S., Staubert 
R. et al. 1997, A\&A, 322, 747
\bibitem[\protect\citeauthoryear{Lebrun et al.}{2003}]{b18} Lebrun F., Leray J.P., Lavocat P. 
et al. 2003 A\&A, 411, 141
\bibitem[\protect\citeauthoryear{Levine et al.}{1984}]{b19} Levine A., Lang F., Lewin W. et al. 
1984, ApJS, 54, 581
\bibitem[\protect\citeauthoryear{Magdziarz \& Zdziarski}{1995}]{b44} Magdziarz P., Zdziarski, A. A. 1995, MNRAS, 273, 837
\bibitem[\protect\citeauthoryear{Maisack et al.}{1993}]{b45} Maisack, M., Mannheim, K., Collmar, W. 1993, A\&A, 319, 397
\bibitem[\protect\citeauthoryear{Malizia et al.}{2005}]{b20} Malizia A., Bassani L., Stephen J.B. et al. 2005, ApJL, 630, 157
\bibitem[\protect\citeauthoryear{Maraschi \& Haardt}{1997}]{b43} Maraschi L. Haardt F. 1997, ASPC, 121, 101
\bibitem[\protect\citeauthoryear{Masetti et al.}{2004}]{b21} Masetti N., Palazzi E., Bassani L. 
et al. 2004, A\&A, 426, 41
\bibitem[\protect\citeauthoryear{Matt}{2001}]{b22} Matt G. 2001, AIPC, 599, 209
\bibitem[\protect\citeauthoryear{Mattson \& Weaver}{2004}]{b23} Mattson B. J., Weaver K. A. 
2004, ApJ, 601, 771
\bibitem[\protect\citeauthoryear{Mukai et al.}{2003}]{b24} Mukai K., Hellier C., Madejski G. et 
al. 2003, ApJ, 597, 479
\bibitem[\protect\citeauthoryear{Mushotzky et al.}{1993}]{b42} Mushotzky, R. F., Done, C., Pounds, K., A. 1993, ARA\&A, 31, 717
\bibitem[\protect\citeauthoryear{Perola et al.}{2002}]{b25} Perola G. C., Matt G., Cappi M. et 
al. 2002, A\&A, 389, 802
\bibitem[\protect\citeauthoryear{Petrucci et al.}{2001}]{b26} Petrucci P. O., Haardt F., 
Maraschi L. et al. 2001, MmSAI, 72, 29
\bibitem[\protect\citeauthoryear{Piro}{1999}]{b27} Piro L. 1999, AN, 320, 236
\bibitem[\protect\citeauthoryear{Reynolds \& Fabian}{1996}]{b28} Reynolds C. S., Fabian, A. C. 
1996, MNRAS, 278, 479
\bibitem[\protect\citeauthoryear{Risaliti}{2002}]{b29} Risaliti G. 2002 A\&A, 386, 379
\bibitem[\protect\citeauthoryear{Sazonov et al.}{2004}]{b30} Sazonov, S.Yu., Revnivtsev, M.G., 
Lutovinov, A.A. et al 2004, A\&A, 421, L21
\bibitem[\protect\citeauthoryear{Sekiguchi \& Menzies}{1990}]{b31} Sekiguchi K., Menzies J. W. 
1990, MNRAS, 245, 66 
\bibitem[\protect\citeauthoryear{Soldi et al.}{2005}]{b32} Soldi S., Beckmann, V., Bassani, L. 
et al. 2005, A\&A, 444, 431
\bibitem[\protect\citeauthoryear{Staubert et al.}{1994}]{b33} Staubert R., K\"onig M., 
Friedrich S. et al. 1994, A\&A, 288, 513 
\bibitem[\protect\citeauthoryear{Terrier et al.}{2003}]{b34} Terrier R., Lebrun F., Bazzano A. 
et al. 2003, A\&A, 411, 167
\bibitem[\protect\citeauthoryear{Ubertini et al.}{2003}]{b35} Ubertini P., Lebrun F., Di Cocco 
G. et al. 2003, A\&A 411, 131
\bibitem[\protect\citeauthoryear{Ubertini et al.}{2005}]{b36} Ubertini P., Bassani L., Malizia 
A. et al. 2005, ApJ Lett., 629, 109
\bibitem[\protect\citeauthoryear{Wilkes et al.}{2001}]{b37} Wilkes B.J., Mathur S., Fiore F. et 
al. 2001, ApJ, 549, 248
\bibitem[\protect\citeauthoryear{Winkler et al.}{2003}]{b38} Winkler C., Courvoisier T., Di 
Cocco G. et al. 2003, A\&A, 411, 1 
\bibitem[\protect\citeauthoryear{Young et al.}{2002}]{b39} Young A.J., Wilson A.S., Terashima 
Y. et al. 2002, ApJ, 564, 176
\bibitem[\protect\citeauthoryear{Zdziarski et al.}{2000}]{b40} Zdziarski A.A., Poutanen J., 
Johnson W.N. 2000, ApJ, 542, 703
\end{thebibliography}
\end{document}